\documentclass{jfm}
\usepackage{graphicx}

\usepackage{epstopdf, epsfig}

\shorttitle{Lagrangian network analysis of turbulent mixing}
\shortauthor{G. Iacobello, S. Scarsoglio, J.G.M. Kuerten, L. Ridolfi}

\title{Lagrangian network analysis \\ of turbulent mixing}

\author{Giovanni Iacobello\aff{1}
  \corresp{\email{giovanni.iacobello@polito.it}},
  Stefania Scarsoglio\aff{1},
  J.G.M. Kuerten\aff{2}, 
 \and Luca Ridolfi\aff{3}}

\affiliation{\aff{1}Department of Mechanical and Aerospace Engineering, Politecnico di Torino, Turin, Italy
\aff{2}Department of Mechanical Engineering, Eindhoven University of Technology, Eindhoven, The Netherlands
\aff{3}Department of Environmental, Land and Infrastructure Engineering, Politecnico di Torino, Turin, Italy}

\begin{document}

\maketitle

\begin{abstract}
		
	A temporal complex network-based approach is proposed as a novel formulation to investigate turbulent mixing from a Lagrangian viewpoint. By exploiting a spatial proximity criterion, the dynamics of a set of fluid particles is geometrized into a time-varying weighted network. Specifically, a numerically solved turbulent channel flow is employed as an exemplifying case. We show that the time-varying network is able to clearly describe the particle swarm dynamics, in a parametrically robust and computationally inexpensive way. The network formalism enables to straightforwardly identify transient and long-term flow regimes, the interplay between turbulent mixing and mean flow advection, and the occurrence of proximity events among particles. Thanks to their versatility and ability to highlight significant flow features, complex networks represent a suitable tool for Lagrangian investigations of turbulence mixing. The present application of complex networks offers a powerful resource for Lagrangian analysis of turbulent flows, thus providing a further step in building bridges between turbulence research and network science.

\end{abstract}

\begin{keywords}
\end{keywords}

\section{Introduction}\label{sec:Introd}
		
		One of the most remarkable features of turbulent flows is their ability to strongly enhance transport and mixing processes. This outstanding property plays a crucial role in many natural phenomena and engineering applications, ranging from chemical reactions and combustion mechanisms \citep[e.g., see][]{warnatz1996combustion, nguyen2018scalar} to biophysics \citep{seuront2004eulerian}, as well as atmospheric dispersion and geophysical phenomena \citep[e.g., see][]{pasquill1983atmospheric, fernando2012handbook}. In spite of many efforts, several issues regarding the understanding and modelling of turbulent mixing -- e.g., anomalous scaling of statistics and signal intermittency -- need to be addressed \citep{warhaft2000passive, sawford2001turbulent, dimotakis2005turbulent, toschi2009lagrangian}. The Lagrangian viewpoint of turbulent mixing -- in which the focus is on particles that are advected by the flow -- has turned out to be better suited than the Eulerian view in many cases \citep{falkovich2001particles, sreenivasan2010lagrangian}. To this aim, the time evolution of pair- and multi-particles has usually been explored in terms of geometrical features -- such as pairwise mean-square separation or multi-particle shape evolution --, and results are typically ensemble averaged \citep[e.g., see][]{salazar2009two, pitton2012anisotropy, bianchi2016evolution, polanco2018relative}. 
		
		In this work, turbulent mixing is investigated in a Lagrangian way by exploiting the recent advances in complex networks, aiming to extend the level of information of classical statistics. Differently to previous approaches, the geometrical representation into complex networks of the particle dynamics offers a twofold tool. On one hand, the network topology systematically geometries the particle dynamics, thus introducing a schematic and synthetic representation of the particle swarm in time. On the other hand, the network formalism provides a robust and well-established framework for studying non-trivial spatio-temporal dynamics of a discrete set of interacting elements \citep{boccaletti2006complex}. Here, we propose a network-based geometrization of particle dynamics in which nodes correspond to fluid particles and links are active by means of a spatial proximity criterion, such that the relative position between particles at any time is enclosed into the network structure. In this way, we obtain a time-varying network which is able to capture in a synthetic way both transient and long-term effects of turbulent mixing, as well as the extent to which particles interact with each other. 
		
	Network science has emerged in last two decades as an effective technique to study real-world complex systems \citep{newman2018networks}, and growing attention is given to the application to turbulent flows \citep[e.g., see][]{charakopoulos2014application, murugesan2015combustion, scarsoglio2016complex, taira2016network, iacobello2018spatial}. Very recently, network-based analysis of fluid flows both in a Eulerian and in a Lagrangian frame has also been carried out. In general, the focus of Lagrangian approaches has been on fluid transport and coherent structure identification, by exploiting the discrete transfer operator \citep{sergiacomi2015flow} and spectral-graph procedures \citep{hadjighasem2016spectral, schlueter2017coherent, padberg2017network, schneide2018probing}. Specifically, the linking criterion is usually based on a similarity measure -- e.g., Euclidean distance, correlation coefficient or Granger causality \citep[see][for an overview]{Donner2017Introduction} -- between particle trajectories evaluated in a given time interval; by doing so, the temporal details of particle trajectories do not explicitly emerge. Differently, in this work the spatio-temporal evolution of particles is captured at each time step, providing a rich and detailed time-dependent picture of turbulent mixing. Therefore, the time-varying network formulation intrinsically highlights the temporal development of particle dynamics due to the turbulent motion.
	
	As a paradigm of possible applications, the proposed approach is shown for a numerically solved turbulent channel flow. In this way, we are able to provide physical insight into the interplay  between the mean flow advection and wall-normal turbulent mixing on particles at different times, thus highlighting the key role of the spatial inhomogeneity.

\section{Flow description and networks building}

	A DNS of a three-dimensional fully-developed incompressible turbulent channel flow was performed at $\Rey_{\tau}=Hu_{\tau}/\nu=950$, where $H$ is half the channel height, $u_{\tau}$ the friction velocity and $\nu$ the kinematic viscosity \citep{kuerten2013lagrangian}. The length of the domain equals $2\pi H$ in streamwise direction, and $\pi H$ in the spanwise direction; in these two directions, periodic boundary conditions are applied to velocity and fluctuating part of the pressure. A constant mean pressure gradient in the streamwise direction is applied to drive the flow. The grid spacing in wall units equals $7.8$ and $3.9$ in the streamwise and spanwise direction, respectively, while the maximum wall-normal grid spacing (at the channel centre) is $7.8$. The total simulation time is $T^+=15200$, while the time-step is $\Delta t^+=4.75$, where the $+$ superscript indicates wall-units normalization. The $T^+$ value was set greater than the wall-normal (Eulerian) mixing time, $T^+_\epsilon$, namely the integral time-scale after which the Taylor dispersion analysis is asymptotically valid \citep{fischer1973longitudinal}. By definition, $T^+_\epsilon \nu/u_\tau^2=H^2/\epsilon_y$ \citep{fernando2012handbook}, where $\epsilon_y=0.067Hu_\tau$ is Elder's vertical mixing coefficient \citep{elder_1959}; as a result, $T^+_\epsilon\approx 14180$.
		
	\begin{figure}
	  \centerline{\includegraphics{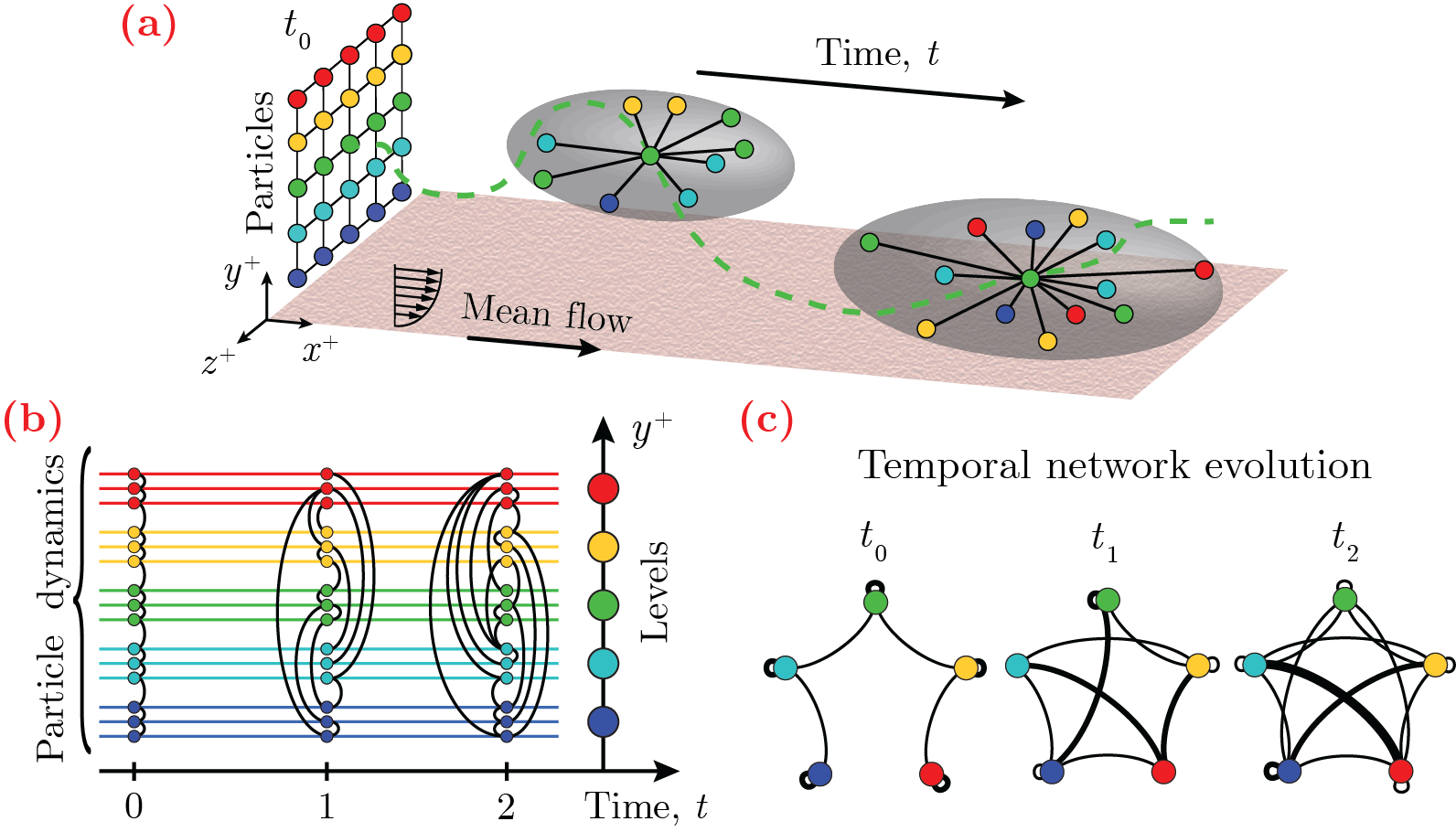}}
	  \caption{(a) Sketch of the setup of the analysis. Particles are initially released from a uniformly spaced grid at $x^+=0$. Coloured spheres represent particles while the ellipsoid is shaded in grey. Connections between particles are illustrated as black lines. (b) Example of particle dynamics depicted as a contact sequence, and (c) temporal evolution of the corresponding networks (the link thickness is proportional to its weight). For all panels, colors from red to blue highlight different starting levels (i.e., different initial $y^+$ values).} \label{fig:methods}
	\end{figure}		
		
		The numerical simulation is run until the statistically stationary condition of the flow is reached. After this condition is fulfilled, particles are seeded in the fully-developed turbulent flow and tracked in time (for more simulation details, see Appendix \ref{appA}). At this initial time, a set of $\left(N_y\times N_z\right)$ fluid particles was arranged as a uniformly distributed grid in the plane $\left(y^+,z^+\right)$ at $x^+=0$, where $(x^+,y^+,z^+)$ are the streamwise, wall-normal and spanwise coordinates, respectively. Although in this work we considered fluid tracers, inertial particles could alternatively be employed. The time-dependent particle positions are then obtained from $\mathrm{d}\boldsymbol{x^+}/\mathrm{d}t^+=\boldsymbol{u}\left(\boldsymbol{x^+}(t^+),t^+\right)$, where $\boldsymbol{x^+}$ is the position of a tracer particle. Since the focus is on the wall-normal mixing process, particles were grouped into $N_y$ wall-normal levels, $l_i$, where each level includes a spanwise-row of fluid particles at the initial time, i.e. $l_i$ depends only on the $y^+$ value of particles at $t^+=0$ (e.g., figure \ref{fig:methods}(a) shows five levels highlighted with different colors). In this work, we set $N_y=N_z=100$, where particles closest to the walls are located at $y^+=9.5$ and the remaining ones are separated by $\Delta y^+=2 \Rey_\tau/N_y=19$, while the initial spanwise separation is $\Delta z^+= 29.85$. The initial particle grouping into levels naturally emerges here as the presence of the walls introduces an inhomogeneous direction, $y^+$. The proposed initial particle arrangement is suitable to numerically represent, for example, a uniformly distributed release of contaminant in the domain. In general, initial particle grouping can be not easily identifiable, resulting in a non-trivial task which requires \textit{ad hoc} partition techniques \citep[e.g., see][]{hadjighasem2017critical, balasuriya2018generalized}.

		To investigate the turbulent dispersion through the network formalism, the interactions between fluid particles were determined based on mutual spatial proximity. Accordingly, if two particles come sufficiently close in space during their motion, a connection is established between them. To specify the particle proximity, we assume that a particle $i$ is connected to a particle $j$ if $i$ lies inside a reference ellipsoid centred at $j$, and \textit{vice versa} (by symmetry). For instance, figure \ref{fig:methods}(a) illustrates the temporal evolution of a particle along its trajectory, and the connections with other particles enclosed in its reference ellipsoid. The ellipsoid was geometrically anchored to each particle location and it was chosen as reference geometry to take into account the anisotropy of the flow. Each ellipsoid is determined by means of its semi-axis lengths, $\boldsymbol{a}=\left( a_x,a_y,a_z\right)$, representing spatial scales of turbulent motion along each Cartesian direction. Accordingly, if the Euclidean distance between a pair of particles is less than $a_i$ in each direction $i=x,y,z$ (namely they are connected), then the two particles share turbulent length scales greater (or equal) than $a_i$. The choice of $\boldsymbol{a}$ is generally a non-trivial task which depends on the specific problem under study, such as the presence of inhomogeneities in the flow or if particles are involved in chemical or biological process (in which specific interactions occur when particles are sufficiently close). When characteristic scales are not known \textit{a priori}, the issue of setting $\boldsymbol{a}$ -- namely to assess the typical length scales in the flow -- can be faced, for instance, by relying on turbulence spectra, correlation functions, as well as coherent structure identification techniques. 
		\\In this work, in order to illustrate the potential of the proposed approach, the semi-axes of the ellipsoid were set proportional to the average pairwise distances in the corresponding Cartesian directions. In this way, the increase of the average mutual distance between particles with time -- that is mainly due to the streamwise dispersion and partly to the spanwise mixing -- is taken into account. By indicating the average Euclidean distance as $\left\langle \boldsymbol{d}\right\rangle$, then $a_i(t^+)=\alpha_i \left\langle d_i(t^+)\right\rangle$, where angular brackets indicate the average over all particle pairs, $i=x,y,z$ and $\boldsymbol{\alpha}=\left(\alpha_x,\alpha_y,\alpha_z\right)$ is a set of parameters. Although $\boldsymbol{\alpha}$ may also explicitly depend on $t^+$, we considered $\alpha_i$ constant in time in order to focus only on the temporal dependency of the semi-axes $\boldsymbol{a}$, as retained in $\left\langle \boldsymbol{d}(t^+)\right\rangle$. In particular, as the simplest case, we selected $\boldsymbol{\alpha}$ as a constant in the range $\alpha\in\left(0,1\right]$, so that the ellipsoid size in each Cartesian direction does not exceed the average distance, $\left\langle \boldsymbol{d}\right\rangle$ (this is crucial along $y^+$ since particles can not exceed the inter-wall height, $2H$). As a reference case, we set $\alpha=0.5$ so that the ellipsoid size in each Cartesian direction is equal to the average distance; however, similar results are obtained with other $\alpha$ values.

		Since particles follow different trajectories due to turbulent motion, the spatial proximity approach results in a non-trivial time-sequence of connections. An example of a geometrical representation of the particle dynamics is shown in figure \ref{fig:methods}(b) as a contact sequence: each connection between pairs of particles (depicted as small coloured circles) is indicated as a black arc, meaning that particles are sufficiently close in space at that time. Therefore, we employed a complex network-based approach to geometrically represent and investigate particle dynamics. Complex networks are defined as graphs -- namely they are made up of entities called nodes which are interconnected by links -- that show non-trivial topological features \citep{boccaletti2006complex, newman2018networks}. In this work, nodes correspond to levels, $l_i$, i.e. spanwise-groups of particles initially at the same $y^+$. By doing so, information of particles starting at the same wall-normal positions is enclosed in each node. Therefore, the metrics extracted from the network increase their statistical significance because they do not represent the dynamics of a single particle. 
		
		For any time, a weight is associated to each link between a pair of nodes $\left(i,j\right)$, which takes into account the total number of connections shared by particles belonging to levels $l_i$ and $l_j$. As a result, particle dynamics is modelled by means of a \textit{time-varying} network \citep{barrat2004weighted, holme2012temporal}, that is a sequence of $N_T=T^+/\Delta t^+$ \textit{weight matrices}, defined as 

		\begin{equation}
			W_{i,j}(\boldsymbol{a},\boldsymbol{\lambda},t^+)=\sum_{p\in l_i}\sum_{q\in l_j}{I_{p,q}(\boldsymbol{a},t^+) K\left(\boldsymbol{\lambda}_{p,q}(t^+)\right)},\qquad i,j=1,...,N_y,
		  \label{eq:W_ijt}
		\end{equation}
		
		\noindent where $W_{i,j}=W_{j,i}$, and the binary indicator function $I_{p,q}$ is equal to $1$ if a particle $p$ lies inside the ellipsoid of a particle $q$ (or \textit{vice versa}) at time $t^+$, and $0$ otherwise. The window function $K$ is a weighting function taking into account the interaction strength, $\boldsymbol{\lambda}_{p,q}=\left(\lambda_x,\lambda_y,\lambda_z\right)_{p,q}$, between particle pairs $\left(p,q\right)$. In this work, $\boldsymbol{\lambda}$ corresponds to the pairwise Euclidean distance (evaluated in wall units) between particles (i.e., $\boldsymbol{\lambda}_{p,q}\equiv\boldsymbol{d}_{p,q}$), but other similarity or distance functions can be adopted. As discussed for the characterization of $\boldsymbol{a}$, the choice of $K$ is conditioned to the specific problem under study and the corresponding aim. For instance, when $K=1$ each link-weight exactly counts the total number of connections shared by particles belonging to pairs of nodes. An example of a time-varying network for $K=1$ is shown in figure \ref{fig:methods}(c): the particle dynamics represented in figure \ref{fig:methods}(b) is geometrized into time-evolving networks, where nodes correspond to levels and thickness of each link is proportional to its weight. 		
		
		Due to its versatility (i.e., by properly setting $\boldsymbol{a}$, $\boldsymbol{\lambda}$ and $K$, as well as the number and the initial arrangement of particles), the proposed time-varying network approach is a powerful tool for turbulence analysis. In fact, the weight matrices, $W_{i,j}(t^+)$, capture and inherit the turbulent mixing information of particles initially located at different $y^+$, at time any $t^+$. In particular, a pair of nodes that is not linked at a given time (i.e., $W_{i,j}=0$) consists of two levels whose particles are not sufficiently close in space (namely, a link is absent in the network topology). On the other hand, non-zero $W_{i,j}$ values quantify the intensity of the connection between levels, namely the extent to which particles are close in space. By doing so, nodes do not vary with time (i.e., they still represent the same levels), while link weights depend on time. 
		\\Finally, it should be noted that the network-based approach recently proposed by \citet{padberg2017network} for the study of Lagrangian transport and mixing (which follows the idea by \citet{Rypina2017}) can be obtained from equation \ref{eq:W_ijt} by setting $\boldsymbol{\lambda}=\boldsymbol{d}$, $a_x=a_y=a_z$, $K=1$ and by checking whether a pair of particles comes sufficiently close in space at least once in the time window considered. Additionally, equation \ref{eq:W_ijt} can be generalized by removing the dependence on $\boldsymbol{a}$ (that is, by setting $a_i=\infty$), thus only keeping the window function $K$ as a similarity measure between particle pairs (e.g., as adopted in Lagrangian coherent structure approaches, see also \citet{hadjighasem2016spectral, schlueter2017coherent, schneide2018probing}).

	\section{Results and discussion}\label{sec:risult}

	In this section, we present the results of the time-varying weighted network approach to investigate turbulent mixing. Two configurations of window function $K$ are explored. 
	\\First, in order to show the main features of the time-varying networks in a simple case, we set $K\equiv K^u=1$, where superscript $u$ indicates a uniform window function (see Sections \ref{subsec:net_struc}-\ref{subsec:charac_mix}); namely, particle interaction is unweighted (binary representation). In this way, each connection is equally weighted, and the connection criterion is only driven by $\boldsymbol{a}$ -- i.e., the main criterion is to check if particles come sufficiently close or not, at any time. In the second configuration (see Section \ref{subsec:K_w}), we set
	\begin{equation}
		K\equiv K^w_{p,q}=\left. {\left(\frac{1}{d_{p,q}+1} - \frac{1}{D_{p,q}+1}\right)} \middle/ {\left(1-\frac{1}{D_{p,q}+1}\right)} \right.,
		\label{eq:K_1d}
	\end{equation}
	
	\noindent where $d_{p,q}$ is the Euclidean distance between particles $p$ and $q$, given that $p$ and $q$ are connected (i.e., $I_{p,q}=1$, see equation \ref{eq:W_ijt}). If the ellipsoid is centred on a particle $p$, $D_{p,q}$ is the distance from $p$ and the point of intersection between the ellipsoid border and the straight line between $p$ and $q$ (the same criterion holds if the ellipsoid is centred on the particle $q$). In this way, the window function is bounded as $K^w\in\left[0,1\right]$, where the ellipsoid border represent an iso-value $K^w=0$, while $K^w=1$ is obtained if particle positions coincide (see also \citet{hadjighasem2016spectral, schneide2018probing} for other works following an inverse distance-based weighting function). In this case, the anisotropy of the flow is explicitly considered as $D_{p,q}$ depends on the length of the ellipsoid semi-axes. By using a monotonically decreasing function for $K$, the smallest spatial scales are more weighted than the largest ones. This corresponds to the assumption that the smallest turbulent scales play a more significant role in particle dynamics.
	
	\begin{figure}
			\centerline{\includegraphics[width=\linewidth]{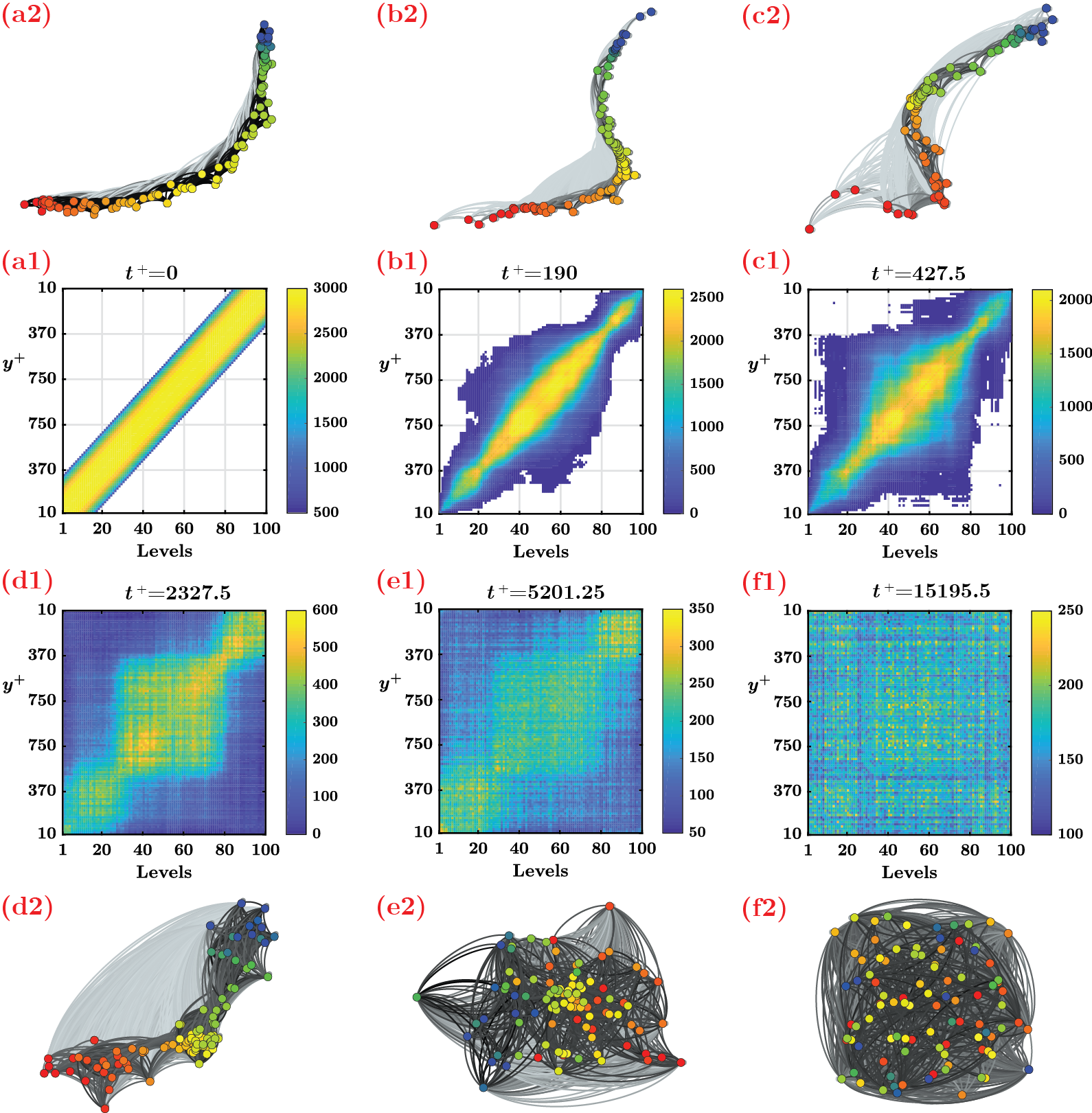}}
			 \caption{Representation of the time-varying network as: (a1-f1) weight matrices, $W_{i,j}$; (a2-f2) their corresponding network topology. In panels (a1-f1), the labels of the matrix ordinates indicate the $y^+$ value (relative to the closest wall, i.e. $y^+ \in \left[ 0,950 \right]$) of the corresponding level reported in the matrix abscissas, while colors represent the weight of the links, $W_{i,j}$. Network visualizations in panels (a2-f2) are obtained through the \textit{OpenOrd} layout algorithm \citep{martin2011openord}; node colors indicate different $y^+$ values and range from red (i.e., level $1$) to blue (i.e., level $100$), while link weights are shown in a grey-scale, where strong links are in black and weak links are in light-grey.}
			\label{fig:W_netw}
		\end{figure}

	\subsection{Network structure of particle dynamics}\label{subsec:net_struc}

		To show the networks from the particle dynamics, in figure \ref{fig:W_netw} the weight matrices at six characteristic times and their corresponding network topology are reported (for further visualizations see \textit{Movie1}). Since particles are initially arranged in a uniform grid at $x^+=0$, each level is simply connected to levels close-in-space in the $\left(y^+,z^+\right)$ plane at $t^+=0$, resulting in the diagonal weight matrix of figure \ref{fig:W_netw}(a1). For small times, the particle dynamics is led by an almost purely advective motion and the particle swarm takes the shape of a bow-like surface (at the very beginning, this surface reproduces the mean velocity profile, $U(y^+)$, in each $\left(x^+,y^+\right)$ plane). Accordingly, the weight matrix exhibits a predominant diagonal pattern, as shown in figure \ref{fig:W_netw}(b1). However, wall-normal turbulent mixing enables out-of-diagonal connections, and an increasing number of weak $W_{i,j}$ values between initially distant levels appears in time (e.g., levels $l_{10}$ and $l_{75}$ in figure \ref{fig:W_netw}(c1)). Sufficiently far enough downstream from $x^+=0$, all levels are interconnected with each other (e.g., see figure \ref{fig:W_netw}(d1)), due to the progressively enhanced transversal mixing. Nevertheless, the mean flow advection is still dominant over mixing at this stage: the initial linear diagonal structure of $W_{i,j}$ evolves into a three-square diagonal pattern, where the central square is bounded between $y^+\approx500$ and $y^+\approx1400$ (namely levels $l_{27}$ and $l_{74}$). This wall-normal coordinate corresponds to the $y^+$ value at which the mean shear, $\partial U/\partial y$, sharply decreases towards zero. In fact, while particles initially located far from the walls experience an almost zero mean shear (thus moving downstream at a high mean velocity), particles close to the wall tend to form two long tails due to the large mean shear close to the wall. The advection process makes the tails progressively stretched along the walls as time increases, highlighting the effect of the mean shear on particle swarm. Therefore, the three-square pattern emerges as a consequence of the mean shear on the particle dynamics. Finally, at some time long after, turbulent mixing becomes as effective as streamwise advection. This is first manifested as a smoothing in the three-square pattern of $W_{i,j}$ (figure \ref{eq:W_ijt}(e1)), and later as a random-like structure (figure \ref{eq:W_ijt}(f1)). This final state represents the Taylor dispersion regime, in which the streamwise particle distribution approaches a Gaussian distribution (see also \textit{Movie2}).

	The temporal characterization of particle dynamics is also highlighted by a different network topology in figure \ref{fig:W_netw}. For short times, the networks show a tree-like elongated structure (figure \ref{fig:W_netw}(a2-c2)) in analogy with the diagonal pattern of the corresponding $W_{i,j}$ (figure \ref{fig:W_netw}(a1-c1)). On the other hand, turbulence mixing -- by enabling links between distant levels -- has the effect to induce a clustered topology, that is a network geometry in which nodes tend to aggregate with each other. In fact, as illustrated in figure \ref{fig:W_netw}(d2), the nodes of the network at $t^+=2327.5$ tend to group based on their wall-normal coordinate: this topology corresponds to the three-square pattern of figure \ref{fig:W_netw}(d1). Finally, for large times the turbulent mixing is much more effective and the network topology develops towards a strongly aggregated pattern with a random-like layout (figure \ref{fig:W_netw}(e2-f2)).
	
\subsection{Advection-mixing regimes and network robustness}\label{subsec:res_parametric}

	\begin{figure}
		\centerline{\includegraphics{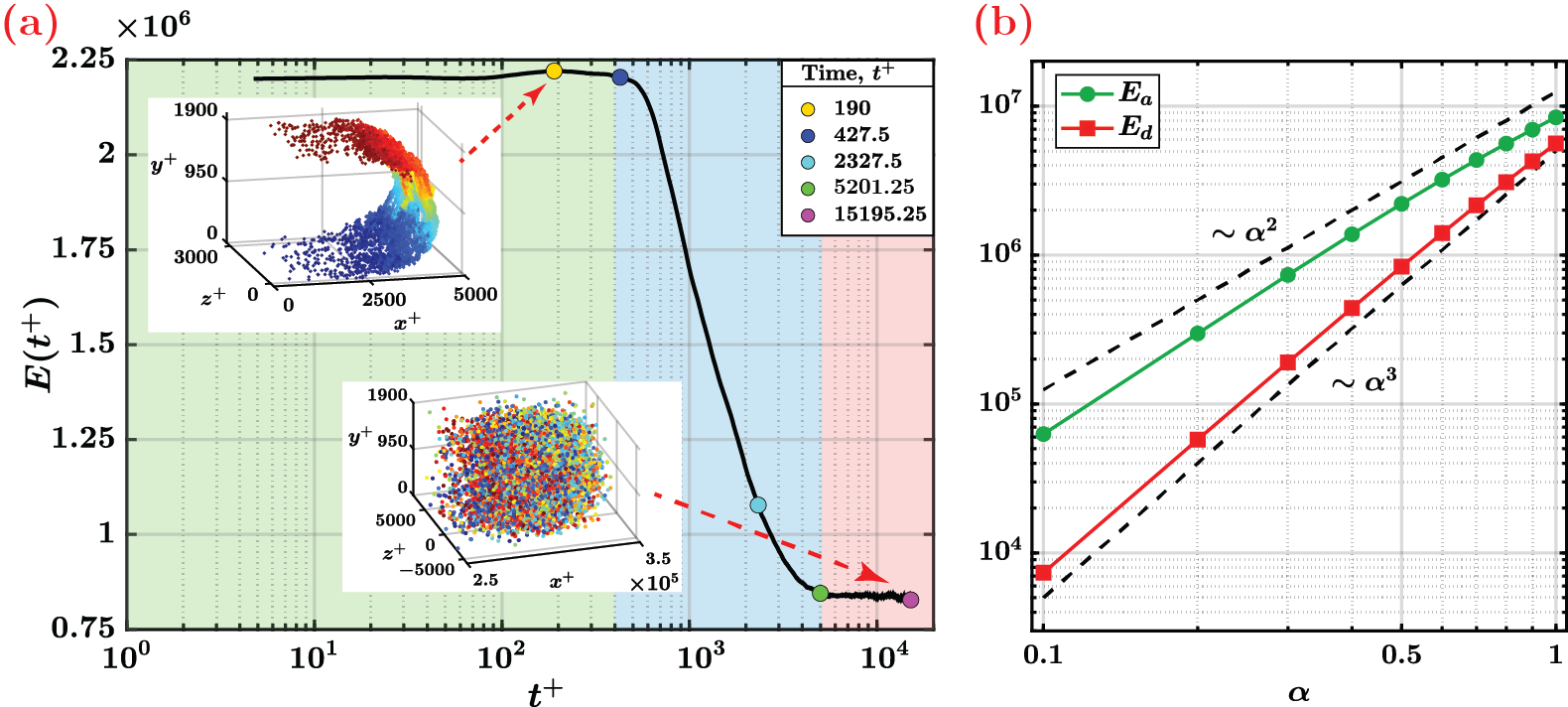}}
		\caption{(a) Total number of connections, $E(t^+)$, as a function of time. Three regimes are highlighted and distinguished by $T^+_a\sim 400$ and $T^+_d\sim 5200$. The two insets show the particle swarms in the first and third regimes, where particle colors indicate their starting level (blue is $l_1$, red is $l_{100}$). The values of $E$ at the same times reported in figure \ref{fig:W_netw} are also illustrated as coloured circles. (b) Effect of $\alpha$ on $E(t^+)$, where $E_a$ and $E_d$ are the average value of $E(t^+)$ over $t^+<T^+_a$ and $t^+>T^+_d$, respectively. Dashed lines indicate the scaling as $\alpha^2$ and $\alpha^3$.} 
		\label{fig:E_sensit}
	\end{figure}
	
	In order to highlight the richness of the information contained in the weighted network as time evolves, we introduce a scalar metric, $E\equiv\sum_i{\sum_j{W_{i,j}}}/2$, that is the total number of network connections established at each time. Figure \ref{fig:E_sensit}(a) shows the behaviour of $E$ as a function of $t^+$, where we see three temporal regimes. The first regime ranges in $t^+\in\left(0,T^+_a\right]$, where $T^+_a\approx 400$ is the time-scale in which particle dynamics is primarily led by streamwise advection. Therefore, in this first regime particles are arranged in a bow-like shape, as shown in the top inset in figure \ref{fig:E_sensit}(a). Since the wall-normal mixing is the main responsible for the activation of distant inter-levels connections, in the first regime the total number of links is almost unchanged and only particles initially close in space are connected with each other (see also figure \ref{fig:W_netw}). As time increases, however, mixing progressively strengthens and $E(t^+)$ decreases up to $t^+\approx 5200$. In fact, if two particles are connected (i.e., each particle lies inside the reference ellipsoid of the other one), wall-normal mixing tends to move the two particles apart in the wall-normal direction. Accordingly, particles tend to deviate from the bow-like profile (typical of the first regime) in the $y^+$ direction, and they come across a region in which the particles are less dense; as a result, in the second regime $E(t^+)$ decreases as turbulent mixing progressively intensifies. Therefore, the second regime is an intermediate stage between an advection-dominant and a mixing-dominant regime. At some time long after, the advection process and the transversal mixing are balanced and the particle dynamics approaches the Taylor asymptotic state. Hence, the third regime is characterized by a nearly constant value of $E(t^+)$, because the particle swarm spreads in the streamwise and spanwise directions without showing any spatial pattern (see the bottom inset in figure \ref{fig:E_sensit}(a)). According to our analysis, this third regime starts at time $T^+_d\approx 5200$, corresponding to the time-scale from which Taylor's dispersion analysis can be applied \citep{fernando2012handbook}. By comparing $T^+_d$ and $T^+_\epsilon$, we found that Taylor's analysis holds for $T^+_d/T^+_\epsilon\gtrsim 0.37$ which is in excellent agreement with the value $0.4$ reported in literature \citep[e.g., see][]{fischer1973longitudinal}. It is worth noting that, while $T^+_\epsilon$ (and $T^+_d\propto T^+_\epsilon$) can be estimated from a dimensional analysis of the conservation equation of a tracer, the value of $T^+_a$ is non-trivial. Indeed, $T^+_a$ is governed by a transient dynamics in which the magnitude of the advection and mixing terms is not easy to quantify. By means of particle geometrization into networks, however, we are able to easily distinguish the onset of both transient (i.e., $T^+_a$) and long-term regimes (i.e., $T^+_d$). It should also be noted that the chosen time resolution ($\Delta t^+=4.75$) is sufficiently accurate for the present analysis, as the network features smoothly evolve over time (see figure \ref{fig:E_sensit}(a)).
	
	\noindent To summarize, the network structures are directly affected over time by wall-normal turbulent mixing or, in general, by the interplay between mixing and advection. In particular, the effect of turbulent mixing on particle dynamics is captured by the total number of connections, $E$, which is able to reveal to which extent wall-normal mixing breaks the initial particle arrangement towards the Taylor asymptotic state.

	For a fixed number of levels, $N_y$, the behaviour of $E(t^+)$ -- and in turn of the weighted network -- basically depends on two modelling parameters: the number of particles in each level, $N_z$, and the constant of proportionality, $\alpha$, between the ellipsoid semi-axes and the average pairwise Euclidean distance. Besides, the DNS spatial resolution could affect the network structure, since a coarse spatial resolution implies an inaccurate velocity field and, in turn, a poor particle position resolution. However, the adopted spatial resolution is sufficient to provide a reliable particle position evaluation \citep{geurts2012ideal,kuerten2013lagrangian}.
	\\To assess the robustness of the proposed approach, we first performed a parametric analysis on $N_z$ while keeping $\alpha=0.5$. By decreasing $N_z$ (keeping $N_y=100$ constant), the total number of particles released in the channel is reduced, thus weakening the statistical significance of the results. However, we found that -- by considering a fraction $1/\delta$ of the total number of particles (with $\delta=2,...,10$) -- the curve of $E(t^+)$ scales down from the reference case shown in figure \ref{fig:E_sensit}(a) proportionally to $1/\delta^2$, as expected, with a relative error below $10\%$. The effect of different $\alpha$ values, instead, is to vary the size of the reference volume for the link activation between particles. Hence, a higher (lower) value of $\alpha$ increases (decreases) the possibility that particles connect with each other. By varying $\alpha$ in the range $\left(0,1\right]$, the curve of $E(t^+)$ is scaled but the values of $T^+_a$ and $T^+_d$ do not change (as for the parametric analysis on $N_z$). However, in this case, $E(t^+)$ does not scale as $\alpha^3$ at any time, as one would expect, but the scaling depends on the regime. In fact, as shown in figure \ref{fig:E_sensit}(b), the mean value of $E(t^+)$ for $t^+<T^+_a$ (namely $E_a$) scales as $\alpha^2$, because in the first regime only a fraction of each ellipsoid is occupied by particles, resulting from the intersection of a bow-like surface with an ellipsoidal volume. Differently, the mean value of $E(t^+)$ for $t^+>T^+_d$ (i.e., $E_d$) scales as $\alpha^3$, because in the third regime particles are spread in all directions and particles occupy the entire volume of each ellipsoid.

\subsection{Characterization of turbulent mixing}\label{subsec:charac_mix}

\begin{figure}
		\centerline{\includegraphics{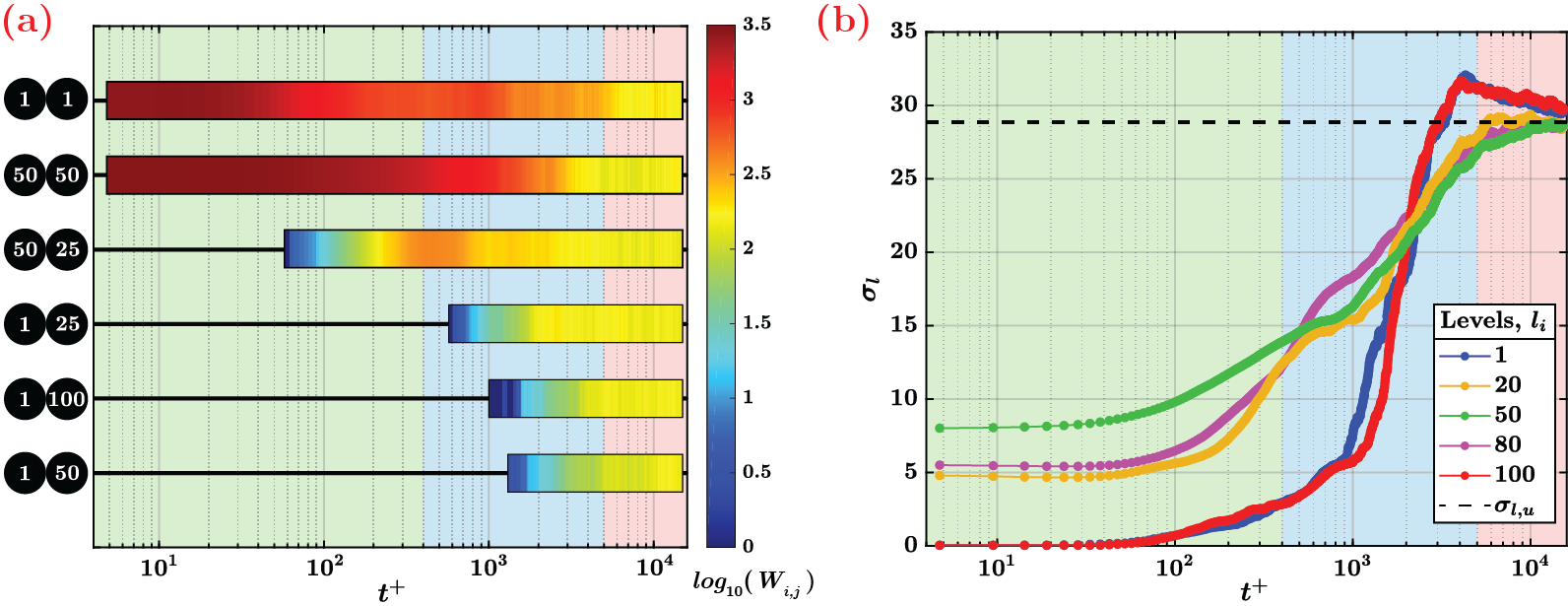}}
	    \caption{(a) Time evolution of $W_{i,j}$ for six pairs of levels. Each horizontal bar corresponds to an entry of the $W_{i,j}$ matrices, while link activation is highlighted by horizontal coloured bands, where color variations indicate the change of the link weight over time proportionally to $log_{10}(W_{i,j})$. (b) Standard deviation, $\sigma_{l_i}$, of different levels inside the ellipsoid of particles in five representative levels. The horizontal dashed line indicates the standard deviation from a discrete uniform distribution in the interval $\left[1,N_y\right]$, namely $\sigma_{l,u}\approx28.87$.}
		\label{fig:wij_stdLvl}
	\end{figure}

	The investigation of $E(t^+)$ provides concise insights into the ensemble behaviour of all levels at each time, and it enables to distinguish different advection-mixing regimes. Nevertheless, from the weight matrices $W_{i,j}$ it is possible to extract much more detailed information. In fact -- as shown in figure \ref{fig:W_netw}(b1-c1) where low $W_{i,j}$ values appear out of the main diagonal -- the key effect of mixing is to promote the activation of links between nodes corresponding to distant levels (e.g., $l_1$ and $l_{100}$). Since particle geometrization into the network framework is based on the spatial proximity, the appearance of a link between two distant levels represents a peculiar event, which is important information, for instance when particles are involved in chemical reactions. In order to characterize such events, in figure \ref{fig:wij_stdLvl}(a) we show the temporal behaviour of $W_{i,j}(t^+)$ for six representative pairs of nodes. If we focus on how connections between particles in the same level change over time (that is the main diagonal of $W_{i,j}$, with $i=j$), link activation starts as expected at the initial time (both for levels close to the wall, $l_1$, and at the center, $l_{50}$) and the weight decreases towards an asymptotic average value. For the level pair $\left(50-25\right)$ -- namely particles initially started at $y^+\simeq940$ and $y^+\simeq465$, respectively --, link activation starts quickly in time, because particles belonging to $l_{25}$ and $l_{50}$ do not experience a strong velocity gradient and turbulent mixing enables their connection after a short time interval. A link between the pair $\left(1-25\right)$, instead, appears only at $t^+\approx 570$, because during the first regime particles in $l_1$ experience higher mean shear than particles in $l_{25}$, so only when the mixing is strong enough a link appears between them. Unexpectedly, as shown in figure \ref{fig:wij_stdLvl}(a), a link between the two furthest levels, $\left(1-100\right)$, appears before a link between levels $\left(1-50\right)$. This can be explained by recalling that if particles in the bow-like swarm are moved apart due to wall-normal mixing, they come across a region less dense of other particles. Therefore, particles in $l_1$ are more likely to connect with particles in $l_{100}$ than particles in level $l_{50}$, because when $W_{1,100}>0$ particles in $l_{50}$ are mainly located far downstream so that $W_{1,50}=0$. Therefore, link activation strongly depends on the flow features, namely both mixing and advection, and it reveals non-trivial results. In particular, since each node represents a set of $N_z$ particles, the activation of a link between two nodes takes into account all the pairwise connections between the $N_z$ particles in each node. By doing so, each link captures and highlights -- through its weight -- ensemble information about the dynamics of the $N_z$ particles in each node. It is worth noting that, for simplicity, we only explored the inter-relation between pairs of nodes, namely $2-$tuples, but the weighted networks comprise the information of all $n-$tuples of nodes.
	
	Finally, we focused on how the presence of different levels inside each ellipsoid varies over time due to the effect of mixing. In fact, any particle belonging to a level $l_i$ is connected -- due to spatial proximity -- with a set of other particles belonging to different levels $l_j$, at each time. This is illustrated in figure \ref{fig:methods}(a) where the presence of different levels inside the ellipsoid of a reference particle is highlighted by different particle colors. To quantify such variability at any time, we evaluated the standard deviation, $\sigma_{l_i}$, of the indices, $j=1,...,N_y$, of levels $l_j$ found inside the ellipsoids of all particles belonging to a reference level, $l_i$. In this way, we quantify the efficiency of mixing between different levels. In figure \ref{fig:wij_stdLvl}(b) we show $\sigma_{l_i}$ as a function of time for five representative levels, $l_i$: due to the progressive strengthening of turbulent mixing, the values of $\sigma_{l_i}$ generally increase with time. Specifically, since in the first regime advection tends to move particles apart in the streamwise direction initially started at different $y^+$, levels close to the walls display lower values of $\sigma_{l_i}$ because they are unlikely to connect with other levels at higher $y^+$ (see top inset in figure \ref{fig:E_sensit}(a)). For long times, instead, all the $\sigma_{l_i}$ approach the value of a uniform distribution, $\sigma_{l,u}$, because of strong turbulent mixing. However, only levels close to the wall (i.e., $l_1$ and $l_100$) show $\sigma_{l_i}$ values above $\sigma_{l,u}$, as a consequence of their preferential connection with very distant levels (as illustrated in figure \ref{fig:wij_stdLvl}(a)). It should also be noted that pairs of levels at a similar distance from the wall (i.e., pairs $1-100$ and $20-80$) show analogous behaviour, since they experience similar dynamics.

\subsection{Time-varying network: weighting connections, $K^w$}\label{subsec:K_w}

\begin{figure}
			\centerline{\includegraphics{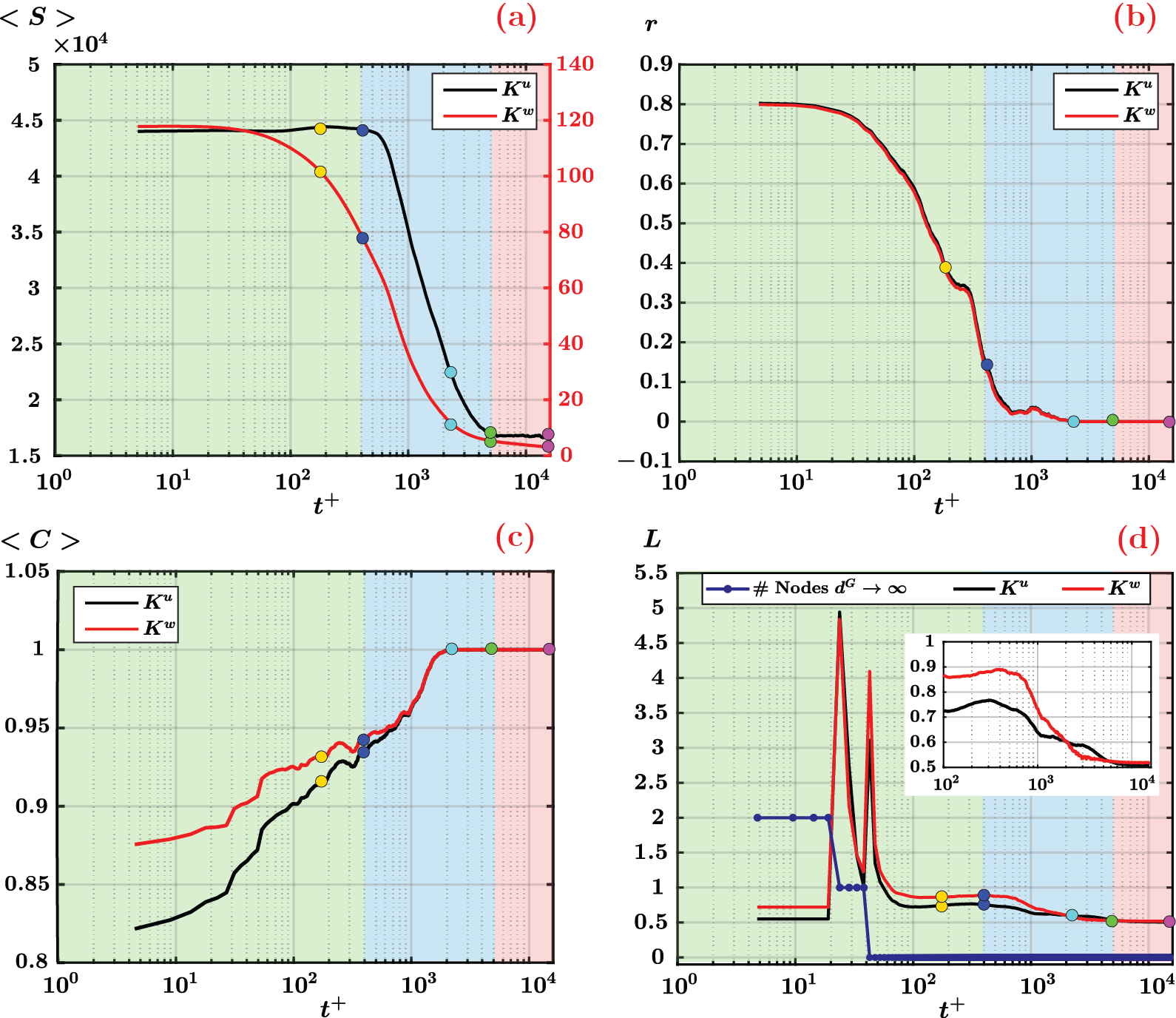}}
			 \caption{Network metrics as a function of time, $t^+$, for the two cases of window function, $K^u$ (black lines) and $K^w$ (red lines). (a) Average strength, $\left\langle S\right\rangle$; the red ordinate axis on the right refers to the range of $\left\langle S\right\rangle$ for the $K^w$ case. (b) Weighted assortativity coefficient, $r$. (c) Average clustering coefficient, $\left\langle C\right\rangle$. (d) Average path length, $L$, and number of disconnected nodes in the networks (blue dotted line). The inset in panel (d) is a zoom of the decreasing behaviour of $L$ with time. In all panels, coloured circles refer to the times reported in figure \ref{fig:W_netw}, while background colors highlight the three advection-mixing regimes (see Section \ref{subsec:res_parametric}).}
			\label{fig:netw_metrics}
		\end{figure}

	The time-varying network built by equally weighting each particle connection (i.e., $K\equiv K^u=1$) shows several remarkable features of the turbulent dispersion over time, as well as its effect on particle dynamics. However, in order to explicitly account for the distance between particles in a particle pair, in the present section we show the results for a non-uniform weighting function, which monotonically decreases with the Euclidean distance, as defined in equation \ref{eq:K_1d}. The ellipsoid is still used as a spatial proximity limit for particle connections -- so that particles outside each ellipsoid are not considered since $I_{p,q}=0$ -- but now particles at smaller spacing inside each ellipsoid are more weighted. It should be noted that the (binary) structure of networks for $K^w$ at any time is the same as shown for $K^u$ in figure \ref{fig:W_netw}. In fact, the diagonal and three-square patterns are also found for $K^w$ in the first and second regime, respectively, while for large times patterns do not emerge. However, by using a non-uniform window function, the values of the link weights (i.e., the link colors in figure \ref{fig:W_netw}) for the $K^w$ case tend to be more intense along the diagonal, as shortest connections are weighted more. In figure \ref{fig:netw_metrics} we show four network metrics to further characterize the time-varying weighted networks in both configurations, $K^u$ and $K^w$. The four metrics are selected in order to progressively highlight the main network features, ranging from a local (i.e., single nodes) to a global (i.e., the entire network) point of view: the node centrality by evaluating the strength; the node pairs by evaluating the assortativity coefficient; the node triples by evaluating the clustering coefficient; and the node $n-$tuples (of higher order) by computing all the shortest paths.
	
	Figure \ref{fig:netw_metrics}(a) shows the average strength, $\left\langle S(t^+)\right\rangle$, of the networks as a function of time (angular brackets indicates average over all nodes). The strength of a node $i$ in a weighted network is defined as $S_i(t^+)\equiv\sum_j{W_{i,j}(t^+)}$, and quantifies the intensity of the relation between node $i$ and all other nodes \citep{barrat2004weighted}. In the case $K^u=1$, $\left\langle S\right\rangle$ is related to the number of particle connections, $E$, as $\left\langle S\right\rangle =2E/N_y $. The behaviour of $\left\langle S(t^+)\right\rangle$ as a function of time for the two cases of $K$ is essentially the same, because (as discussed in section \ref{subsec:res_parametric}) the effect of turbulent dispersion on particle dynamics is fully captured by the network structure corresponding to $K^u=1$. However, the values of $\left\langle S(t^+)\right\rangle$ for the case $K^u$ are globally higher (about two order of magnitude) than the corresponding values for $K^w$ (note that in figure \ref{fig:netw_metrics}(a) there are two ordinate axes), as $K^w$ ranges in the interval $\left[0,1\right]$. The difference in the temporal behaviour between the two $K$ configurations is more evident in the first regime, where $\left\langle S\right\rangle$ for the $K^w$ case (red curve in figure \ref{fig:netw_metrics}(a)) decreases more rapidly. In the first regime, advection is dominant over mixing and most of the links are present between nodes initially close in space (i.e., along the diagonal of $W_{i,j}$). However, as mentioned in Section \ref{subsec:res_parametric}, wall-normal turbulent mixing promotes the activation of links between initially distant nodes, as emerges from the appearance of out-of-diagonal links in figure \ref{fig:W_netw}(a1-c1) and the increase of standard deviation in figure \ref{fig:wij_stdLvl}(b). These out-of-diagonal links correspond to connections between particles that are distant in space within an ellipsoid, thus corresponding to very low $K^w$ values. In fact, if a particle $q$ enters inside the ellipsoid centred in a particle $p$ at a given time (with $p$ and $q$ belonging to distant initial levels), it is very likely that $q$ is close to the border of the ellipsoid of $p$ (where $K^w_{p,q}=0$). On the other hand, for the $K^u$ case, connections between particles either close or distant in space are equally weighted. Therefore, the activation of distant links (i.e., out-of-diagonal points in figure 2(b1-c1)) makes the values of $\left\langle S\right\rangle$ decrease faster for the $K^w$ case than for the $K^u$ case.
	\\Following the concept of strength, in figure \ref{fig:netw_metrics}(b), we show the assortativity coefficient, $r(t^+)$, which is defined as the Pearson correlation coefficient of the strength of the nodes at the ends of each link \citep{boccaletti2006complex}. A network is named assortative ($0< r\leq 1$) or disassortative ($-1\leq r< 0$) if nodes tend to link with other similar or dissimilar nodes (this similarity is here measured through the strength), respectively; otherwise the network is said non-assortative (i.e., $r\approx 0$). In both $K$ configurations, $r(t^+)$ similarly decreases from approximatively $0.8$ at small times to zero at large times. In the first regime (especially at short times) nodes are primarily linked with other similar nodes (e.g., see figure \ref{fig:W_netw}(a1-c1)), which implies that the networks are assortative; on the other hand, for long times, the networks lose any pattern (e.g., see figure \ref{fig:W_netw}(e1-f1)), thus showing a non-assortative behaviour.

		Figure \ref{fig:netw_metrics}(c) shows the average clustering coefficient, $\left\langle C(t^+)\right\rangle$, for the configurations $K^u$ (black curve) and $K^w$ (red curve). The clustering coefficient quantifies the probability that two randomly chosen neighbours of a node $i$ (i.e., two nodes linked to $i$) are also neighbours, thus ranging in the interval $\left[0,1\right]$. It is formally defined as $C_i=N_\Delta(i)/N_3(i)$, where $N_\Delta(i)$ is the number of triangles and $N_3(i)$ is the number of triples involving node $i$, respectively \citep{newman2018networks}. A weighted clustering coefficient takes into account the interaction intensity between nodes comprising triplets \citep{barrat2004weighted}. As shown in figure \ref{fig:netw_metrics}(c), $\left\langle C\right\rangle$ increases to one for both configurations as time increases: indeed, for large times particles are well mixed with each other and it is very likely that nodes form triangles, namely nodes are locally very well inter-connected. The effect of a weighted network is to enhance the local cohesiveness especially in the first regime, thus the values of $\left\langle C\right\rangle$ are the highest in the $K^w$ case, for which spatial proximity plays a more significant role.
	
	Another feature of complex network structure is the concept of shortest path, namely the path with minimal cost between two nodes, where a path is an alternating sequence of nodes (and edges) considered only once \citep{boccaletti2006complex}. We investigate the average path length, $L(t^+)$, that is the average of all shortest path lengths in the network. Specifically, the average path length is $L=\sum_{i\neq j}{d^G_{i,j}}/(N_y^2-N_y)$, where $d^G_{i,j}$ is the length (cost) of the shortest path between node $i$ and $j$ in a graph $G$ \citep{boccaletti2006complex}. In this work, since the weight associated to each link, $W_{i,j}$, represents the intensity of spatial proximity between nodes, higher $W_{i,j}$ implies closer distances. Accordingly, the shortest paths for the evaluation of $L(t^+)$ are computed by using a weighting cost $\left\langle W\right\rangle/W_{i,j}$, where $\left\langle W\right\rangle$ is the average link weight of the networks at any time \citep{opsahl2010node}. The entries of the weighted matrix, $W_{i,j}$, are normalized through $\left\langle W\right\rangle$ in order to consider the change of $W_{i,j}$ range as time increases (e.g., see the colorbar ranges in figure \ref{fig:W_netw}(a1-f1)). Therefore, a high weight means a low cost of the path. Figure \ref{fig:netw_metrics}(d) shows the behaviour of $L$ as a function of $t^+$. The networks during the first regime are able to display high $L$ values, since two nodes are reachable via a high-cost path. By inspecting the values of $d^G_{i,j}$ we found that outliers in $L$ for $t^+\in\left[ 20,50\right]$ are due to the nodes starting very close to the walls (namely, $l_i$ with $i=\left\lbrace 1,100\right\rbrace$). In the first regime, nodes starting very close to the walls are disconnected from the other nodes in the network (see blue-dotted curve in figure \ref{fig:netw_metrics}(d)) -- namely $d^G_{1,j}=d^G_{100,j}=\infty$ -- because of a strong effect of mean shear on particle positions (see also figure \ref{fig:W_netw}(b1) or the visualizations in \textit{Movie1}). Since disconnected nodes are conventionally excluded from the computation of the average path length \citep{boccaletti2006complex}, the peaks in $L$ appear when nodes $1$ and $100$ are linked to the other nodes, that is at $t^+\approx 25$ and $t^+\approx 40$, respectively. From the point of view of the network topology, the re-attachment of the disconnected nodes implies high-cost shortest paths, as the values of $d^G_{i,j}$ for nodes $i=\left\lbrace 1,100\right\rbrace$ are two orders of magnitude larger than $d^G_{i,j}$ for all the other nodes. As a consequence, very high values of $L$ are detected. It should be noted that, although the effect of the mean shear is evident in the first regime, disconnected nodes are not expected for $t^+\rightarrow 0$, as follows from the initial particle arrangement. For long times, instead, $L$ tends to decrease as an effect of turbulent mixing. By comparing the two weighted configurations, $K^u$ and $K^w$, the values of $L$ in the first two regimes are higher for the $K^w$ case because (as mentioned for the assortativity and the clustering coefficient) the window function highlights the local mixing by enhancing the spatial proximity. In the third regime, instead, both $K$ configurations approach the same constant $L$ value.
	
	In conclusion, the analysis of the metrics at different network levels (from the single node, to node pairs and triples, as well as the shortest path of variable length) of the time-varying weighted network, is able to unveil both the main general features and the presence of extreme cases in the particle dynamics.

\section{Conclusions}\label{sec:conclusion}
	
		The proposed Lagrangian network-based approach is exemplified by means of DNS of a turbulent channel flow, where the dynamics of fluid particles is characterized by a spatial proximity criterion. The resulting time-varying weighted network is fully able to inherit the non-trivial time sequence of connections between (groups of) particles, which emerges due to turbulent motion. Indeed, we can identify in a straightforward way the characteristic regimes of particle dynamics, the appearance of peculiar events (e.g., the time of first contact between initially distant levels), as well as the intensity of wall-normal turbulent mixing (quantified by the total number of links). Accordingly, the potential of the Lagrangian-based networks is twofold, since the time-varying weighted network -- represented by the weight matrices --  captures in a unique framework both the qualitative spatial features of the particle swarm and the strength of turbulent mixing. The proposed complex network geometrization reveals to be robust and frame-invariant (as the Euclidean distance is used). Moreover, this approach is computationally affordable for a typical number of tracers of the order of $10^4 - 10^5$, and it is thus suitable for experimental techniques, such as particle tracking velocimetry, where the number of tracers is usually of the order of $10^2 - 10^3$ \citep{kim2013vortex}. Due to its versatility, particle geometrical representation into time-varying networks can easily be extended to other flows and other tracers, such as inertial particles or passive scalars. Based on present findings, Lagrangian-based networks can pave the way for a systematic network-based investigation of turbulent mixing, especially in the context of dispersion modelling.

\section*{Acknowledgements}

	This work was sponsored by NWO Exacte en Natuurwetenschappen (Physical Sciences) for the use of supercomputer facilities, with financial support from the Netherlands Organization for Scientific Research, NWO.

\appendix

\section{Numerical simulation details}\label{appA}

	A DNS of a three-dimensional fully-developed incompressible turbulent channel flow was performed at $\Rey_\tau=950$. To this end, the continuity equation and the Navier-Stokes (NS) equations
	
	\begin{equation}
		\nabla\cdot{\boldsymbol{u}}=0,\qquad \frac{\partial{\boldsymbol{u}}}{\partial t}+\frac{1}{\rho}\nabla p={\boldsymbol{f}}-{\boldsymbol{\omega}}\times{\boldsymbol{u}}+\nu\nabla^2{\boldsymbol{u}},
		\label{eq:cont_NS}
	\end{equation}
	
	\noindent are solved. In these equations $\rho$ and $\boldsymbol{u}$ denote mass density and velocity of the fluid, $p$ the total pressure, $\boldsymbol{\omega}$ the vorticity and $t$ is time. The friction Reynolds number of the flow is kept fixed by prescribing the mean driving force per unit mass, ${\boldsymbol{f}}$, in the streamwise direction parallel to the plates. The velocity, $\boldsymbol{u}$, satisfies no-slip conditions at the two plates, i.e. $y^+=0$ and $y^+=1900$. In the other two directions, $x^+$ and $z^+$, periodic boundary conditions are applied for the velocity and the fluctuating part of the pressure. In the two periodic directions of the domain a Fourier-Galerkin approach is applied, and in the wall-normal direction $y^+$ a Chebyshev-tau method. The continuity equation is exactly satisfied by using the wall-normal component of the vorticity vector and the Laplacian of the wall-normal velocity component as dependent variables, instead of the three velocity components. The non-linear terms in the Navier-Stokes equations are calculated in the physical space by fast Fourier transform, with application of the $3/2$ rule in both periodic directions. The equations are integrated in time by a combination of a three-stage explicit Runge-Kutta method and the Crank-Nicolson method. In the two periodic directions $768$ Fourier modes are used, while in the wall-normal direction $385$ Chebyshev polynomials are employed. The time step used in the simulation equals $0.095\nu u_{\tau}^{-2}$, while the grid spacing are $\Delta x^+=7.8$, $\Delta y^+_{max}\approx 7.8$ (at the channel centre), and $\Delta z^+=3.9$.
		
	 In order to extract particle trajectories, $\boldsymbol{x^+}(t^+)$, the equation $\mathrm{d}\boldsymbol{x^+}/\mathrm{d}t^+=\boldsymbol{u}\left(\boldsymbol{x^+}(t^+),t^+\right)$, is solved with the same explicit second-order accurate Runge-Kutta method as used in the solution of the NS equations. The fluid velocity is interpolated to the particle location by tri-linear interpolation. The accuracy of the numerical method has been assessed in previous papers \citep{geurts2012ideal,kuerten2013lagrangian}.

\bibliographystyle{jfm}
\bibliography{biblio_Iacobello_et_al}

\begin{thebibliography}{40}
\expandafter\ifx\csname natexlab\endcsname\relax\def\natexlab#1{#1}\fi
\def\au#1{#1} \def\ed#1{#1} \def\yr#1{#1}\def\at#1{#1}\def\jt#1{\textit{#1}}
  \def\bt#1{#1}\def\bvol#1{\textbf{#1}} \def\vol#1{#1} \def\pg#1{#1}
  \def\publ#1{#1}\def\arxiv#1{#1}\def\org#1{#1}\def\st#1{\textit{#1}}

\bibitem[Balasuriya {\em et~al.\/}(2018)Balasuriya, Ouellette \&
  Rypina]{balasuriya2018generalized}
{\sc \au{Balasuriya, S.}, \au{Ouellette, N.~T.} \& \au{Rypina, I.~I.}}
  \yr{2018}  \at{Generalized lagrangian coherent structures}.  \jt{Physica D} .

\bibitem[Barrat {\em et~al.\/}(2004)Barrat, Barth{\'e}lemy \&
  Vespignani]{barrat2004weighted}
{\sc \au{Barrat, A.}, \au{Barth{\'e}lemy, M.} \& \au{Vespignani, A.}} \yr{2004}
   \at{Weighted evolving networks: coupling topology and weight dynamics}.
  \jt{Phys. Rev. Lett.}  \bvol{92}~(22),  \pg{228701}.

\bibitem[Bianchi {\em et~al.\/}(2016)Bianchi, Biferale, Celani \&
  Cencini]{bianchi2016evolution}
{\sc \au{Bianchi, S.}, \au{Biferale, L.}, \au{Celani, A.} \& \au{Cencini, M.}}
  \yr{2016}  \at{On the evolution of particle-puffs in turbulence}.  \jt{Eur.
  J. Mech. B-Fluids}  \bvol{55},  \pg{324--329}.

\bibitem[Boccaletti {\em et~al.\/}(2006)Boccaletti, Latora, Moreno, Chavez \&
  Hwang]{boccaletti2006complex}
{\sc \au{Boccaletti, S.}, \au{Latora, V.}, \au{Moreno, Y.}, \au{Chavez, M.} \&
  \au{Hwang, D.U.}} \yr{2006}  \at{Complex networks: {S}tructure and dynamics}.
   \jt{Phys. Rep.}  \bvol{424}~(4-5),  \pg{175--308}.

\bibitem[Charakopoulos {\em et~al.\/}(2014)Charakopoulos, Karakasidis,
  Papanicolaou \& Liakopoulos]{charakopoulos2014application}
{\sc \au{Charakopoulos, A.K.}, \au{Karakasidis, T.E.}, \au{Papanicolaou, P.N.}
  \& \au{Liakopoulos, A.}} \yr{2014}  \at{The application of complex network
  time series analysis in turbulent heated jets}.  \jt{Chaos}  \bvol{24}~(2),
  \pg{024408}.

\bibitem[Dimotakis(2005)]{dimotakis2005turbulent}
{\sc \au{Dimotakis, P.~E.}} \yr{2005}  \at{Turbulent mixing}.  \jt{Annu. Rev.
  Fluid Mech.}  \bvol{37},  \pg{329--356}.

\bibitem[Donner {\em et~al.\/}(2017)Donner, Hern{\'a}ndez-Garc{\'\i}a \&
  Ser-Giacomi]{Donner2017Introduction}
{\sc \au{Donner, R.~V.}, \au{Hern{\'a}ndez-Garc{\'\i}a, E.} \& \au{Ser-Giacomi,
  E.}} \yr{2017}  \at{Introduction to focus issue: Complex network perspectives
  on flow systems}.  \jt{Chaos}  \bvol{27}~(3),  \pg{035601}.

\bibitem[Elder(1959)]{elder_1959}
{\sc \au{Elder, J.~W.}} \yr{1959}  \at{The dispersion of marked fluid in
  turbulent shear flow}.  \jt{J. Fluid Mech.}  \bvol{5}~(4),  \pg{544–560}.

\bibitem[Falkovich {\em et~al.\/}(2001)Falkovich, Gawȩdzki \&
  Vergassola]{falkovich2001particles}
{\sc \au{Falkovich, G.}, \au{Gawȩdzki, K.} \& \au{Vergassola, M.}} \yr{2001}
  \at{Particles and fields in fluid turbulence}.  \jt{Rev. Mod. Phys.}
  \bvol{73}~(4),  \pg{913}.

\bibitem[Fernando(2012)]{fernando2012handbook}
{\sc \au{Fernando, H. J.~S.}} \yr{2012} {\em Handbook of Environmental Fluid
  Dynamics, Volume Two: Systems, Pollution, Modeling, and Measurements\/}.
  \publ{CRC press}.

\bibitem[Fischer(1973)]{fischer1973longitudinal}
{\sc \au{Fischer, H.~B.}} \yr{1973}  \at{Longitudinal dispersion and turbulent
  mixing in open-channel flow}.  \jt{Annu. Rev. Fluid Mech.}  \bvol{5}~(1),
  \pg{59--78}.

\bibitem[Geurts \& Kuerten(2012)]{geurts2012ideal}
{\sc \au{Geurts, B.~J.} \& \au{Kuerten, J.G.M.}} \yr{2012}  \at{Ideal
  stochastic forcing for the motion of particles in large-eddy simulation
  extracted from direct numerical simulation of turbulent channel flow}.
  \jt{Phys. Fluids}  \bvol{24}~(8),  \pg{081702}.

\bibitem[Hadjighasem {\em et~al.\/}(2017)Hadjighasem, Farazmand, Blazevski,
  Froyland \& Haller]{hadjighasem2017critical}
{\sc \au{Hadjighasem, A.}, \au{Farazmand, M.}, \au{Blazevski, D.},
  \au{Froyland, G.} \& \au{Haller, G.}} \yr{2017}  \at{A critical comparison of
  lagrangian methods for coherent structure detection}.  \jt{Chaos}
  \bvol{27}~(5),  \pg{053104}.

\bibitem[Hadjighasem {\em et~al.\/}(2016)Hadjighasem, Karrasch, Teramoto \&
  Haller]{hadjighasem2016spectral}
{\sc \au{Hadjighasem, A.}, \au{Karrasch, D.}, \au{Teramoto, H.} \& \au{Haller,
  G.}} \yr{2016}  \at{Spectral-clustering approach to lagrangian vortex
  detection}.  \jt{Phys. Rev. E}  \bvol{93}~(6),  \pg{063107}.

\bibitem[Holme \& Saram{\"a}ki(2012)]{holme2012temporal}
{\sc \au{Holme, P.} \& \au{Saram{\"a}ki, J.}} \yr{2012}  \at{Temporal
  networks}.  \jt{Phys. Rep.}  \bvol{519}~(3),  \pg{97--125}.

\bibitem[Iacobello {\em et~al.\/}(2018)Iacobello, Scarsoglio, Kuerten \&
  Ridolfi]{iacobello2018spatial}
{\sc \au{Iacobello, G.}, \au{Scarsoglio, S.}, \au{Kuerten, J.G.M.} \&
  \au{Ridolfi, L.}} \yr{2018}  \at{Spatial characterization of turbulent
  channel flow via complex networks}.  \jt{Phys. Rev. E}  \bvol{98}~(1),
  \pg{013107}.

\bibitem[Kim {\em et~al.\/}(2013)Kim, Hussain \& Gharib]{kim2013vortex}
{\sc \au{Kim, D.}, \au{Hussain, F.} \& \au{Gharib, M.}} \yr{2013}  \at{Vortex
  dynamics of clapping plates}.  \jt{J. Fluid Mech.}  \bvol{714},  \pg{5--23}.

\bibitem[Kuerten \& Brouwers(2013)]{kuerten2013lagrangian}
{\sc \au{Kuerten, J.G.M.} \& \au{Brouwers, J.J.H.}} \yr{2013}  \at{Lagrangian
  statistics of turbulent channel flow at {R}e$_\tau$= 950 calculated with
  direct numerical simulation and {L}angevin models}.  \jt{Phys. Fluids}
  \bvol{25}~(10),  \pg{105108}.

\bibitem[Martin {\em et~al.\/}(2011)Martin, Brown, Klavans \&
  Boyack]{martin2011openord}
{\sc \au{Martin, S.}, \au{Brown, W.~M.}, \au{Klavans, R.} \& \au{Boyack,
  K.~W.}} \yr{2011} Open{O}rd: an open-source toolbox for large graph layout.
  \bt{In {\em Visualization and Data Analysis\/}}, ,  \vol{vol. 7868},  \pg{p.
  786806}.

\bibitem[Murugesan \& Sujith(2015)]{murugesan2015combustion}
{\sc \au{Murugesan, M.} \& \au{Sujith, R.I.}} \yr{2015}  \at{Combustion noise
  is scale-free: transition from scale-free to order at the onset of
  thermoacoustic instability}.  \jt{J. Fluid Mech.}  \bvol{772},
  \pg{225--245}.

\bibitem[Newman(2018)]{newman2018networks}
{\sc \au{Newman, M.}} \yr{2018} {\em Networks\/}, 2nd edn.  \publ{Oxford
  University Press}.

\bibitem[Nguyen \& Papavassiliou(2018)]{nguyen2018scalar}
{\sc \au{Nguyen, Q.} \& \au{Papavassiliou, D.~V.}} \yr{2018}  \at{Scalar mixing
  in anisotropic turbulent flow}.  \jt{AIChE Journal}  \bvol{64}~(7),
  \pg{2803--2815}.

\bibitem[Opsahl {\em et~al.\/}(2010)Opsahl, Agneessens \&
  Skvoretz]{opsahl2010node}
{\sc \au{Opsahl, T.}, \au{Agneessens, F.} \& \au{Skvoretz, J.}} \yr{2010}
  \at{Node centrality in weighted networks: Generalizing degree and shortest
  paths}.  \jt{Soc. Networks}  \bvol{32}~(3),  \pg{245--251}.

\bibitem[Padberg-Gehle \& Schneide(2017)]{padberg2017network}
{\sc \au{Padberg-Gehle, K.} \& \au{Schneide, C.}} \yr{2017}  \at{Network-based
  study of {L}agrangian transport and mixing}.  \jt{Nonlinear Proc. Geoph.}
  \bvol{24}~(4),  \pg{661}.

\bibitem[Pasquill \& Smith(1983)]{pasquill1983atmospheric}
{\sc \au{Pasquill, F.} \& \au{Smith, F.B.}} \yr{1983}  \at{Atmospheric
  diffusion: study of the dispersion of windborne material from industrial and
  other sources.}  \jt{John Wiley \& Sons} .

\bibitem[Pitton {\em et~al.\/}(2012)Pitton, Marchioli, Lavezzo, Soldati \&
  Toschi]{pitton2012anisotropy}
{\sc \au{Pitton, E.}, \au{Marchioli, C.}, \au{Lavezzo, V.}, \au{Soldati, A.} \&
  \au{Toschi, F.}} \yr{2012}  \at{Anisotropy in pair dispersion of inertial
  particles in turbulent channel flow}.  \jt{Phys. Fluids}  \bvol{24}~(7),
  \pg{073305}.

\bibitem[Polanco {\em et~al.\/}(2018)Polanco, Vinkovic, Stelzenmuller, Mordant
  \& Bourgoin]{polanco2018relative}
{\sc \au{Polanco, J.~I.}, \au{Vinkovic, I.}, \au{Stelzenmuller, N.},
  \au{Mordant, N.} \& \au{Bourgoin, M.}} \yr{2018}  \at{Relative dispersion of
  particle pairs in turbulent channel flow}.  \jt{Int. J. Heat Fluid Fl.}
  \bvol{71},  \pg{231--245}.

\bibitem[Rypina \& Pratt(2017)]{Rypina2017}
{\sc \au{Rypina, I.~I.} \& \au{Pratt, L.~J.}} \yr{2017}  \at{Trajectory
  encounter volume as a diagnostic of mixing potential in fluid flows}.
  \jt{Nonlinear Proc. Geoph.}  \bvol{24}~(2),  \pg{189--202}.

\bibitem[Salazar \& Collins(2009)]{salazar2009two}
{\sc \au{Salazar, J.~P.} \& \au{Collins, L.~R.}} \yr{2009}  \at{Two-particle
  dispersion in isotropic turbulent flows}.  \jt{Annu. Rev. Fluid Mech.}
  \bvol{41},  \pg{405--432}.

\bibitem[Sawford(2001)]{sawford2001turbulent}
{\sc \au{Sawford, B.}} \yr{2001}  \at{Turbulent relative dispersion}.
  \jt{Annu. Rev. Fluid Mech.}  \bvol{33}~(1),  \pg{289--317}.

\bibitem[Scarsoglio {\em et~al.\/}(2016)Scarsoglio, Iacobello \&
  Ridolfi]{scarsoglio2016complex}
{\sc \au{Scarsoglio, S.}, \au{Iacobello, G.} \& \au{Ridolfi, L.}} \yr{2016}
  \at{Complex networks unveiling spatial patterns in turbulence}.  \jt{Int. J.
  Bifurc. Chaos}  \bvol{26}~(13),  \pg{1650223}.

\bibitem[Schlueter-Kuck \& Dabiri(2017)]{schlueter2017coherent}
{\sc \au{Schlueter-Kuck, K.~L.} \& \au{Dabiri, J.~O.}} \yr{2017}  \at{Coherent
  structure colouring: {I}dentification of coherent structures from sparse data
  using graph theory}.  \jt{J. Fluid Mech.}  \bvol{811},  \pg{468--486}.

\bibitem[Schneide {\em et~al.\/}(2018)Schneide, Pandey, Padberg-Gehle \&
  Schumacher]{schneide2018probing}
{\sc \au{Schneide, C.}, \au{Pandey, A.}, \au{Padberg-Gehle, K.} \&
  \au{Schumacher, J.}} \yr{2018}  \at{Probing turbulent superstructures in
  {R}ayleigh-{B}{\'e}nard convection by {L}agrangian trajectory clusters}.
  \jt{Phys. Rev. Fluids}  \bvol{3}~(11),  \pg{113501}.

\bibitem[Ser-Giacomi {\em et~al.\/}(2015)Ser-Giacomi, Rossi, L{\'o}pez \&
  Hernandez-Garcia]{sergiacomi2015flow}
{\sc \au{Ser-Giacomi, E.}, \au{Rossi, V.}, \au{L{\'o}pez, C.} \&
  \au{Hernandez-Garcia, E.}} \yr{2015}  \at{Flow networks: {A} characterization
  of geophysical fluid transport}.  \jt{Chaos}  \bvol{25}~(3),  \pg{036404}.

\bibitem[Seuront \& Schmitt(2004)]{seuront2004eulerian}
{\sc \au{Seuront, L.} \& \au{Schmitt, F.~G.}} \yr{2004}  \at{Eulerian and
  {L}agrangian properties of biophysical intermittency in the ocean}.
  \jt{Geophys. Res. Lett.}  \bvol{31}~(3).

\bibitem[Sreenivasan \& Schumacher(2010)]{sreenivasan2010lagrangian}
{\sc \au{Sreenivasan, K.~R.} \& \au{Schumacher, J.}} \yr{2010}  \at{Lagrangian
  views on turbulent mixing of passive scalars}.  \jt{Philos. T. R. Soc. A}
  \bvol{368}~(1916),  \pg{1561--1577}.

\bibitem[Taira {\em et~al.\/}(2016)Taira, Nair \& Brunton]{taira2016network}
{\sc \au{Taira, K.}, \au{Nair, A.~G.} \& \au{Brunton, S.~L.}} \yr{2016}
  \at{Network structure of two-dimensional decaying isotropic turbulence}.
  \jt{J. Fluid Mech.}  \bvol{795}.

\bibitem[Toschi \& Bodenschatz(2009)]{toschi2009lagrangian}
{\sc \au{Toschi, F.} \& \au{Bodenschatz, E.}} \yr{2009}  \at{Lagrangian
  properties of particles in turbulence}.  \jt{Annu. Rev. Fluid Mech.}
  \bvol{41},  \pg{375--404}.

\bibitem[Warhaft(2000)]{warhaft2000passive}
{\sc \au{Warhaft, Z.}} \yr{2000}  \at{Passive scalars in turbulent flows}.
  \jt{Annu. Rev. Fluid Mech.}  \bvol{32}~(1),  \pg{203--240}.

\bibitem[Warnatz {\em et~al.\/}(1996)Warnatz, Maas \&
  Dibble]{warnatz1996combustion}
{\sc \au{Warnatz, J.}, \au{Maas, U.} \& \au{Dibble, R.~W.}} \yr{1996} {\em
  Combustion: physical and chemical fundamentals, modeling and simulation,
  experiments, pollutant formation\/}.  \publ{Springer}.

\end{thebibliography}

\end{document}